# Linear Algebraic Approaches to Neuroimaging Data Compression: A Comparative Analysis of Matrix and Tensor Decomposition Methods for High-Dimensional Medical Images

Jaeho Kim, Daniel David, and Ana Vizitiv

**ABSTRACT**

This paper evaluates Tucker decomposition and Singular Value Decomposition (SVD) for compressing neuroimaging data. Tucker decomposition preserves multi-dimensional relationships, achieving superior reconstruction fidelity and perceptual similarity. SVD excels in extreme compression but sacrifices fidelity. The results highlight Tucker decomposition's suitability for applications requiring the preservation of structural and temporal relationships.

## 1 INTRODUCTION

**Research Question**

This paper aims to answer the following question: "How can matrix and tensor decomposition methods be applied to compress high-dimensional neuroimaging data, optimizing storage efficiency while preserving critical image quality?"

**Background / Motivation**

Image compression is essential in modern data management, particularly in medical imaging, where data quality directly impacts diagnostic accuracy. While traditional compression methods have been effective for general-purpose images, high-dimensional neuroimaging data poses unique challenges, requiring advanced mathematical approaches. Linear algebraic methods, with their ability to capture complex data relationships and reduce dimensionality while preserving essential information, present promising solutions.

Despite progress in managing neuroimaging data, many existing techniques struggle to balance compression efficiency and image fidelity, particularly for multi-dimensional datasets. Advanced linear algebraic methods, such as matrix and tensor decompositions, remain underutilized in medical image compression. Tensors, which generalize scalars, vectors, and matrices to higher dimensions, provide a natural framework for representing multi-dimensional neuroimaging data. These methods, such as Higher-Order Singular Value Decomposition (HOSVD), are effective in capturing the intricate spatial and temporal relationships in neuroimaging datasets, enabling significant dimensionality reduction (Hackbusch, 2012; Kolda & Bader, 2009; Zhou et al., 2013; Cichocki et al., 2016).

Functional Magnetic Resonance Imaging (fMRI) exemplifies the challenges of managing large neuroimaging datasets. fMRI data, represented as three-dimensional (3D) volumetric images or four-dimensional (4D) time-series datasets, are inherently large and complex, presenting significant storage and management challenges (Smith, 2004). For instance, large-scale studies like the NeuroGrid Stroke Exemplar trial generated approximately 12,000 scans, highlighting the financial and logistical difficulties of storing, sharing, and analyzing such extensive data (Van Horn & Toga, 2014; Wardlaw et al., 2007).

Advancements in neuroimaging, such as increased spatial and temporal resolutions, have further escalated dataset sizes. Early diffusion MRI

(dMRI) studies captured diffusion along six directions, while modern approaches now resolve over 512 directions, significantly expanding data volume (Van Horn & Toga, 2014; Wardlaw et al., 2007). With neuroimaging dataset sizes doubling approximately every 26 months since 1995 and exceeding 20GB per study by 2015, efficient compression methods are essential (Van Horn & Toga, 2014).

Traditional matrix-based approaches and recent tensor-based methods offer potential solutions for managing these growing datasets. Identifying the most effective techniques is crucial to ensure that the exponential growth of neuroimaging data does not hinder its clinical and experimental applications (Calhoun & Sui, 2016; Dinov, 2016; Poldrack & Farah, 2015). This project aims to evaluate and compare linear algebraic compression techniques, particularly singular value decomposition (SVD) and Tucker decomposition, through their demonstrated use on a sample fMRI dataset; thereby providing insights into their applications for optimizing neuroimaging data management.

## 2 MATRIX-BASED COMPRESSION

SVD is a powerful method for compressing high-dimensional data while retaining essential features. By breaking down a matrix into its core components, SVD identifies the most significant patterns and structures within the data (Golub & Van Loan, 2013). SVD decomposes a matrix $A \in R^{M \times N}$ into three components:

$$A = U\Sigma V^T$$

Here, $U \in R^{M \times N}$ and $V \in R^{N \times N}$ are orthogonal matrices representing the left and right singular vectors, respectively, and $\Sigma \in R^{M \times N}$ is a diagonal matrix containing the singular values in descending order. These singular values quantify the importance of each corresponding singular vector (Strang, 2019).

To compress an image, we follow three steps:

1. <u>Decomposition</u>: Apply SVD to the image matrix $A$, obtaining $U$, $\Sigma$, and $V^T$.
2. <u>Truncation</u>: Retain only the top $k$ singular values and their associated singular vectors, where $k$ is chosen such that the retained singular values account for a specified percentage (e.g., 90-99%) of the total variance. This step removes less significant components that contribute minimally to the image's structure.
3. <u>Reconstruction</u>: Use the truncated components to reconstruct an approximation of the original matrix:

$$A_K = U_K \Sigma_K V_K^T$$

Here, $U_K$, $\Sigma_K$, $V_K^T$ are the truncated versions of $U$, $\Sigma$, and $V$, respectively.

This process achieves significant compression while maintaining the image's perceptual quality. For example, retaining just 50 singular values out of 500 can reduce storage requirements by up to 90% while preserving most visual details.

SVD-based compression is particularly effective for image data because it captures global patterns, such as edges and textures, which are essential for human perception. Additionally, it is computationally efficient and can be applied to other data types, such as audio and video, making it a versatile tool for dimensionality reduction and data compression (Eckart & Young, 1936).

## 3 TENSOR-BASED COMPRESSION

Tensor decomposition provides a way to analyze high-dimensional neuroimaging data by working with multiple dimensions at once. While matrix methods must flatten multi-dimensional data, tensor approaches can keep the original data structure intact (Kolda & Bader, 2009).

In our study of fMRI data, we work with data that naturally has four dimensions. Each brain scan captures spatial information through the height, width, and depth of the brain, creating a three-dimensional volume. When multiple scans are taken over time to track brain activity, this creates a fourth dimension. This temporal aspect is crucial as it shows how brain activity patterns change throughout the scanning session.

We represent this data as a fourth-order tensor because each dimension captures important information that we need to preserve. The three spatial dimensions together show the brain's structure in 3D space, while the time dimension shows how brain activity changes during the scan. Using a fourth-order tensor lets us analyze all these aspects together rather than having to break them apart (Zhou et al., 2013).

The Tucker decomposition expresses this fourth-order tensor $X \in \mathbb{R}^{I_1 \times I_2 \times I_3 \times I_4}$ as:

$$X \approx G \times_1 U_1 \times_2 U_2 \times_3 U_3 \times_4 U_4$$

Here, **G** is called the core tensor and contains the main patterns in the data, while $U_1$ through $U_4$ are matrices that help reconstruct the original data (Cichocki et al., 2016). These matrices represent patterns in each dimension: $U_1$, $U_2$, and $U_3$ capture spatial patterns, and $U_4$ captures temporal patterns.

To compress an image, we follow three steps:

1. <u>Decomposition</u>: Decompose the data using the formula above.
2. <u>Truncation</u>: Remove less important components that contribute less than 1% to the total data variation.
3. <u>Reconstruction</u>: Reconstruct the data using the remaining components.

This approach allows us to significantly reduce file sizes while keeping the features doctors need for diagnosis. This method works particularly well for brain imaging data because it maintains the relationships between different parts of the brain and how they change over time. Studies show it can significantly reduce data size while preserving the important medical details needed for analysis (Jonmohamadi et al., 2020).

## 4 METHOD

**Sample Data: BOLD5000 fMRI Dataset**

The raw fMRI data used in this study is derived from the BOLD5000 dataset, which provides volumetric brain scans of participants as they viewed a diverse set of 5,254 images from visual datasets such as COCO, ImageNet, and SUN (Chang et al., 2019). The data was collected using blood oxygenation level-dependent (BOLD) imaging, a functional MRI technique that measures brain activity by detecting changes in blood flow. Each scan captures 3D spatial volumes over time, forming a 4D time-series dataset, making it highly relevant for exploring the challenges of high-dimensional neuroimaging data compression.

This dataset was chosen for its complexity and real-world applicability, as it exemplifies the significant storage and processing demands of modern neuroimaging studies. Where the required computing power was too great, we used data from the first participant's first session as a representative sample of the dataset. This approach was justified due to the standardized nature of the data according to BIDS (Brain Imaging Data Structure) standards, a widely adopted framework designed to ensure uniformity and reproducibility in neuroimaging data.

The BIDS standards define a clear organization for data and metadata, including file naming conventions, metadata specifications, and directory structures, facilitating consistent preprocessing and analysis across studies (Gorgolewski et al., 2016). This standardization enhances data interoperability and ensures that results can be reliably replicated and extended, even when using subsets of the dataset. Additionally, the controlled environment and procedures during data collection further support the sample's representativeness.

Preprocessing steps, further detailed in the sections below, involved extracting 3D volumetric scans, applying noise and slice timing corrections, and restructuring the data into matrix and or tensor form to enable SVD analysis and Tucker decomposition.

In this study, we implement SVD and Tucker decomposition to analyze and compress neuroimaging data stored in the DICOM (.dcm)

format, a widely used standard for storing and transmitting medical imaging data. The DICOM format facilitates the integration of images with associated metadata, such as patient information, imaging parameters, and spatial resolution, ensuring interoperability across medical devices and systems (Bidgood et al, 1997). Each .dcm file contains a three-dimensional brain scan captured at a specific time point, with multiple .dcm files combining to form a four-dimensional time series dataset (Bidgood Jr & Horii, 1992; Cox et al., 2004).

The large size and diverse content of the BOLD5000 dataset provided an ideal test case for evaluating how SVD and Tucker decomposition can balance storage efficiency and image quality preservation, directly addressing the research question posed in this study.

**Matrix-Based Compression**

In this part we will analyze how the image compression using SVD influences images in quality and size.

The following code processes a given image according to different values of *r*:

```python
U, S, VT = np.linalg.svd(X,full_matrices=False)
S = np.diag(S)

j = 0
for r in (5, 10, 20, 30, 40, 50, 75, 100):
    # Construct approximate image
    Xapprox = U[:,:r] @ S[0:r,:r] @ VT[:r,:]
    plt.figure(j+1)
    j += 1
    img = plt.imshow(Xapprox)
    img.set_cmap('gray')
    plt.axis('off')
    plt.title('r = ' + str(r))
    plt.show()
```

**Figure 1.** Python code for compressing an image using SVD. The code processes the image for different rank values *r* (5, 10, 20, 30, 40, 50, 75, and 100) to construct approximations.

Then, we process the value of Memory size, MSE, PSNR according to each r value:

```python
import os
# Save the images with different r values
for r in (5, 10, 20, 30, 40, 50, 75, 100):
    Xapprox = U[:, :r] @ S[0:r, :r] @ VT[:r, :]
    plt.figure()
    img = plt.imshow(Xapprox)
    img.set_cmap('gray')
    plt.axis('off')
    plt.title('r = ' + str(r))
    filename = f'approx_image_r_{r}.png'
    plt.savefig(filename, bbox_inches='tight', pad_inches=0)
    plt.close()
    print(f'Size of {filename}: {os.path.getsize(filename)} bytes')
```

```python
from skimage.metrics import mean_squared_error, peak_signal_noise_ratio

for r in (5, 10, 20, 30, 40, 50, 75, 100):
    Xapprox = U[:, :r] @ S[0:r, :r] @ VT[:r, :]
    mse = mean_squared_error(X, Xapprox)
    psnr = peak_signal_noise_ratio(X, Xapprox, data_range=X.max() - X.min())
    print(f'For r = {r}, MSE: {mse}, PSNR: {psnr} dB')

print('\nMSE values range from 0-∞, with lower being better.')
print('PSNR values range from 20-50 dB, with higher being better.')
```

**Figure 2.** Python code for evaluating the impact of rank r on image compression. This snippet computes memory size, MSE, and PSNR for reconstructed images at the specified ranks *r*.

MSE (Mean Squared Error) is a metric that quantifies the average squared difference between the pixel values of the original image and the reconstructed (compressed) image. It measures the reconstruction error, with lower MSE values indicating a closer match to the original image. It is defined as:

$$\text{MSE} = \frac{1}{m \times n} \sum_{i=1}^{m} \sum_{j=1}^{n} \left( I(i,j) - \hat{I}(i,j) \right)^2$$

where *I(i,j)* is the pixel value of the original image, *Î(i,j)* is the pixel value of the reconstructed image, and *m*, *n* are the image dimensions.

PSNR (Peak Signal-to-Noise Ratio) is a measure of image quality based on the logarithmic ratio of the peak signal (maximum possible pixel intensity) to the noise (distortion). It is expressed in decibels (dB), with higher PSNR values indicating better image quality. It is calculated as:

$$\text{PSNR} = 10 \cdot \log_{10} \left( \frac{MAX_I^2}{\text{MSE}} \right)$$

where $MAX_I$ is the maximum pixel intensity value (e.g., 255 for 8-bit images).

We hypothesize that increasing the rank r in the SVD method will improve image quality at the cost of larger file sizes. As r increases, the compressed image size grows due to the inclusion of additional singular values, which preserve more details of the original image. This improvement in quality is reflected by a decrease in MSE and an increase in PSNR, demonstrating the trade-off between compression efficiency and reconstruction fidelity.

## Tensor-Based Compression

```python
def apply_tucker_decomposition(tensor, ranks):
    # Convert to tensorly tensor
    tl_tensor = tl.tensor(tensor)
    print(f"Tensor shape before decomposition: {tl_tensor.shape}")

    # Perform Tucker decomposition
    core, factors = tucker(tl_tensor, rank=ranks)
    reconstructed = tl.tucker_to_tensor((core, factors))
    return core, factors, reconstructed
```

**Figure 3.** Implementation of Tucker decomposition using TensorLy library. The function converts input data to a tensor, performs Tucker decomposition with variable ranks, and returns a core tensor, factor matrices, and reconstructed tensor.

The Tucker decomposition framework (Cichocki et al., 2016; Kolda & Bader, 2009) decomposes the 4D tensor $X \in \mathbb{R}^{I1 \times I2 \times I3 \times I4}$ into a core tensor and factor matrices, expressed as:

$$X \approx G \times_1 U_1 \times_2 U_2 \times_3 U_3 \times_4 U_4$$

Here, $G$ represents the core tensor containing the interactions between components, while $U_n$ denotes the factor matrices representing the principal components in each mode. This decomposition effectively captures both spatial relationships within individual brain scans and temporal patterns across the time series.

Implementation begins with converting the sequential .dcm files into a unified tensor structure. For tensor decomposition, the rank array is defined as $[r_1, r_2, r_3]$, where $r_1$, $r_2$, and $r_3$ denote the ranks for the number of slices, width dimension, and height dimension, respectively. Optimal rank selection for each mode utilizes relative reconstruction error analysis, which evaluates the difference between original and reconstructed tensors using RMSE and SSIM metrics (Cichocki et al., 2016). This process balances compression efficiency against data fidelity. The compression process retains components that contribute more than 1% to the total explained variance as measured by singular values (Zhou et al., 2013), with thresholds determined through empirical analysis of reconstruction quality.

To assess reconstruction accuracy, we employ Root Mean Square Error (RMSE) (Cichocki et al., 2016; Wang & Ahuja, 2005), calculated as:

$$\text{RMSE} = \frac{||X - \hat{X}||_F}{||X||_F}$$

where $X$ represents the original data, $\hat{X}$ represents the reconstructed data, $||\cdot||_F$ represents the Frobenius norm, which calculates the sum of the squares of all entries in a tensor and then takes the square root. RMSE thereby provides a numerical measure of how much information is lost in the compression process.

In addition, we employ the Structural Similarity Index Measure (SSIM), calculated as:

$$\text{SSIM}(X, \hat{X}) = \frac{(2\mu_X \mu_{\hat{X}} + C1)(2\sigma_{X\hat{X}} + C2)}{(\mu_X^2 + \mu_{\hat{X}}^2 + C1)(\sigma_X^2 + \sigma_{\hat{X}}^2 + C2)}$$

where $\mu_X$ and $\mu_{\hat{X}}$ are mean intensities, $\sigma^2_X$ and $\sigma^2_{\hat{X}}$ are variances, $\sigma_{X\hat{X}}$ is covariance, and $C_1$, $C_2$ are small constants for stability. SSIM evaluates luminance, contrast, and structural similarity to ensure clinically relevant features are preserved. Collectively, these employments ensure the preservation of clinically relevant image features, such as anatomical structures (Jack & Holtzman, 2013), tissue contrast (Filippi & Rocca, 2011), lesions (Lansberg et al, 2012), and functional activity (Fox & Raichle, 2007).

Compression efficiency is measured through compression ratio analysis and computational resource utilization (Van Horn & Toga, 2014; Smith, 2004). The compression ratio (CR) is calculated as:

$$\text{CR} = \frac{\text{Original Data Size}}{\text{Compressed Data Size}}$$

This metric quantifies the extent of data reduction achieved, with higher values indicating more efficient compression. Computational resource utilization is assessed by measuring processing time and memory usage during compression and reconstruction, providing insights into the practical feasibility of the method.

We hypothesize that tensor-based methods, such as Tucker decomposition, will outperform matrix-based methods like SVD in preserving the structural and temporal integrity of high-dimensional neuroimaging data. By retaining multi-dimensional relationships, tensor approaches are expected to achieve better compression ratios and lower reconstruction errors, as measured by RMSE and SSIM, while effectively balancing compression efficiency and data fidelity.

The validation framework incorporates cross-validation to ensure generalizability. Metrics such as RSME and SSIM are used to evaluate the reconstruction quality, with RMSE quantifying numerical accuracy and SSIM assessing perceptual similarity and the preservation of the aforementioned diagnostically significant features in the compressed data.

## 5 RESULTS & ANALYSIS

### Matrix-Based Compression

The analysis was conducted using approximations of a sample image (in gray scales) at varying ranks *r*, with the following results:

| Rank (r) | Image Size (bytes) | MSE | PSNR (dB) |
|---|---|---|---|
| 5 | 166,957 | 7,748.91 | 29.11 |
| 10 | 175,032 | 5,516.69 | 30.59 |
| 20 | 186,320 | 3,651.88 | 32.38 |
| 30 | 204,479 | 2,786.70 | 33.55 |
| 40 | 216,747 | 2,262.66 | 34.46 |
| 50 | 219,578 | 1,891.03 | 35.24 |
| 75 | 228,968 | 1,281.17 | 36.93 |
| 100 | 234,826 | 915.85 | 38.39 |

**Table 1.** The relationship between rank, image size, MSE, and PSNR following SVD analysis. Refer to **Figure 4** in the Appendix for a plot detailing their relationship.

As *r* increases, the size of the compressed image grows, demonstrating a direct relationship between rank and memory usage. For instance, at r = 5, the size was 166,957 bytes, while at r = 100, it increased to 234,826 bytes - a 40.6% growth.

Moreover, higher ranks consistently produced lower MSE values and higher PSNR values, indicating improved image quality. For instance, at r = 5, MSE was 7,748.91, and PSNR was 29.11 dB. At r = 100, MSE dropped to 915.85, and PSNR increased to 38.39 dB. These trends confirm the hypothesis that retaining more singular values improves image reconstruction accuracy.

Combined, these results affirm the hypothesis that increasing r leads to better image quality at the cost of larger file sizes. This demonstrates the utility of SVD in balancing compression and quality for specific applications.

### Tensor-Based Compression

Applying Tucker decomposition to the neuroimaging dataset demonstrates a clear trade-off between compression efficiency, measured by the compression ratio, and reconstruction fidelity, quantified using RMSE and SSIM. The yielded results are presented in **Table 2** and **Figure 5** in the Appendix.

Lower rank configurations, such as [5, 5, 5], achieved extreme compression with a compression ratio of 506.72, reducing the dataset size to 3,836 bytes from the original size of 1,944,474 bytes. However, this configuration resulted in high reconstruction error (RMSE of 103.83) and low perceptual similarity (SSIM of 0.51). As the rank increased, RMSE decreased, and SSIM improved significantly. For instance, the [30, 30, 30] configuration achieved a compression ratio of 21.88 while maintaining acceptable reconstruction fidelity with an RMSE of 47.20 and SSIM of 0.81. Higher ranks, such as [50, 50, 50], reduced RMSE to 29.54 and improved SSIM to 0.90, with a compression ratio of 5.58. At the highest tested rank, [75, 75, 75], Tucker decomposition yielded an RMSE of 11.86 and an SSIM of 0.98, with a compression ratio of 1.90.

For comparison, SVD was applied to the same neuroimaging dataset, with ranks ranging from 5 to 100. At rank 5, SVD achieved a compressed size of 166,957 bytes and a mean squared error (MSE) of 7,748.91. As the rank increased, the compressed size increased, while the MSE decreased. At rank 30, SVD produced a compressed size of 204,479 bytes with an MSE of 2,786.70. At the highest rank tested, rank 100, the compressed size was 234,826 bytes, with an MSE of 915.85. These results indicate that SVD achieves significant compression but at the cost of a higher MSE compared to Tucker decomposition for equivalent compression ratios.

The trade-offs between compression efficiency and reconstruction fidelity for both methods are visualized in **Figure 5**. Tucker decomposition provides better perceptual similarity (SSIM) and structural preservation for the same compression ratio, making it more suitable for tasks requiring high-quality reconstructions, such as clinical neuroimaging. For example, at a compression ratio of approximately 5.5, Tucker decomposition achieved an RMSE of 29.54 with an SSIM of 0.90, whereas SVD produced a compressed size of 219,578 bytes and a higher MSE of 1,891.03 (equivalent to an RMSE of approximately 43.48, assuming RMSE = sqrt(MSE)).

The "sweet spot" for Tucker decomposition lies at rank configurations of [40, 40, 40] and [50, 50, 50], which balance acceptable reconstruction fidelity (RMSE of 37.59 and 29.54, respectively) with reasonable compression ratios (10.30 and 5.58, respectively). For SVD, the optimal trade-off occurs at rank 30, where the compressed size is 204,479 bytes, and the MSE is 2,786.70 (RMSE approximately 52.78). However, SVD does not explicitly retain multi-dimensional relationships, which limits its performance in preserving perceptual similarity compared to Tucker decomposition.

These results support the hypothesis that tensor-based methods, such as Tucker decomposition, outperform matrix-based methods like SVD in preserving the structural and temporal integrity of high-dimensional neuroimaging data. Tucker decomposition consistently achieved lower RMSE values compared to SVD at similar compression ratios. This indicates that Tucker decomposition's ability to retain multi-dimensional relationships gives it a clear advantage in preserving data fidelity while balancing compression efficiency. While SVD offers more straightforward dimensionality reduction, it lacks the structural preservation inherent in Tucker decomposition, as evidenced by its higher reconstruction errors. Thus, Tucker decomposition aligns well with the hypothesis and is a more effective method for neuroimaging data compression in scenarios requiring structural and temporal integrity.

# 6    CONCLUSION

This study demonstrates the efficacy of linear algebraic approaches, particularly Tucker decomposition and SVD, in compressing high-dimensional neuroimaging data. By leveraging these methods, we evaluated the trade-offs between compression efficiency, as measured by compression ratio, and reconstruction fidelity, quantified through RMSE.

Our results reveal that Tucker decomposition excels in preserving the structural and temporal integrity of neuroimaging data by retaining multi-dimensional relationships. It consistently achieved lower RMSE values compared to SVD for similar compression ratios, making it a more effective method for applications where perceptual and structural fidelity are critical. For example, Tucker decomposition at a compression ratio of 5.58 achieved an RMSE of 29.54, whereas SVD resulted in higher RMSE overall. These findings confirm the hypothesis that tensor-based methods like Tucker decomposition outperform matrix-based approaches like SVD in preserving the essential characteristics of neuroimaging datasets.

This work highlights the importance of selecting the appropriate method and rank configuration based on the specific requirements of an application. Tucker decomposition is recommended for clinical and research scenarios where preserving structural and temporal relationships is paramount. In contrast, SVD may be suitable for tasks where computational efficiency and storage constraints take precedence, owing to its comparatively lower computational complexity.

Future work should explore hybrid approaches combining the strengths of both methods, as well as extensions to incorporate additional constraints, such as real-time processing requirements or domain-specific features. By addressing these considerations, we aim to further advance the utility of linear algebraic techniques in managing the growing demands of neuroimaging data.

# APPENDIX

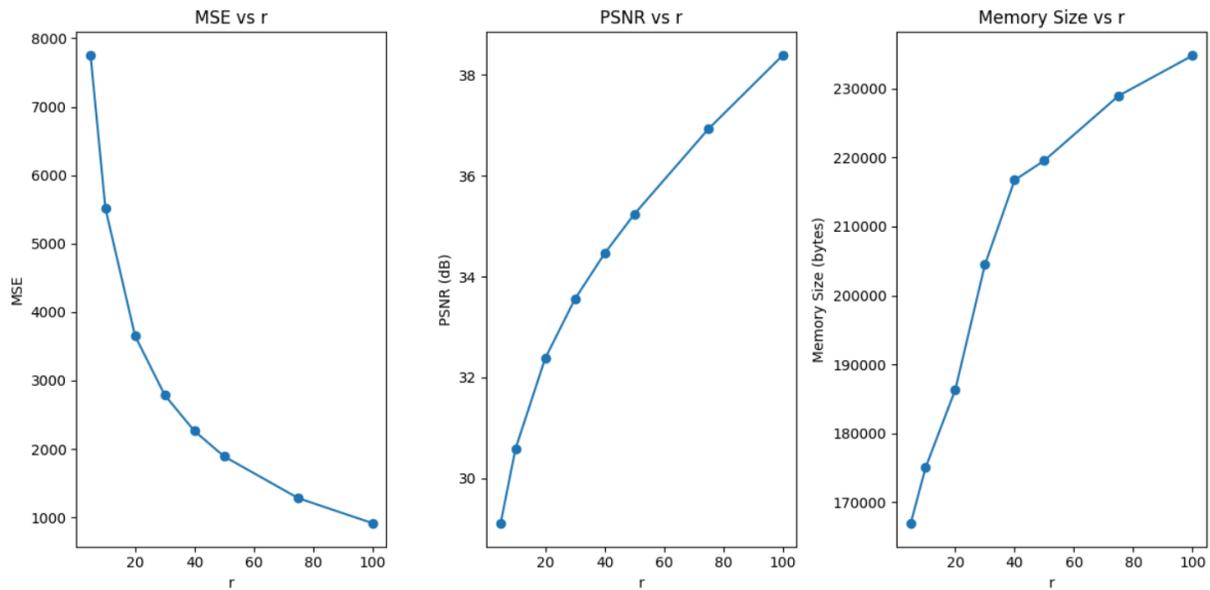

**Figure 4.** Relationship between rank *r*, image size, mean squared error (MSE), and peak signal-to-noise ratio (PSNR) following Singular Value Decomposition (SVD) analysis. The plots illustrate how increasing the rank *r* enhances image reconstruction quality (lower MSE and higher PSNR) at the cost of increased size.

| Method | Rank | Compressed Size (bytes) | Compression Ratio | RMSE | SSIM |
|---|---|---|---|---|---|
| Tucker | [5, 5, 5] | 3,836 | 506.72 | 103.83 | 0.51 |
| Tucker | [10, 10, 10] | 9,559 | 203.49 | 80.73 | 0.60 |
| Tucker | [20, 20, 20] | 34,161 | 56.92 | 59.37 | 0.73 |
| Tucker | [30, 30, 30] | 88,891 | 21.88 | 47.20 | 0.81 |
| Tucker | [40, 40, 40] | 188,701 | 10.30 | 37.59 | 0.87 |
| Tucker | [50, 50, 50] | 348,470 | 5.58 | 29.54 | 0.90 |
| Tucker | [75, 75, 75] | 1,025,942 | 1.90 | 11.86 | 0.98 |
| SVD | rank 5 | 166,957 | 11.64 | 87.97 | N/A |
| SVD | rank 10 | 175,032 | 11.11 | 74.26 | N/A |
| SVD | rank 20 | 186,320 | 10.43 | 60.42 | N/A |
| SVD | rank 30 | 204,479 | 9.51 | 52.78 | N/A |
| SVD | rank 50 | 219,578 | 8.85 | 43.48 | N/A |
| SVD | rank 75 | 228,968 | 8.49 | 35.81 | N/A |
| SVD | rank 100 | 234,826 | 8.28 | 30.27 | N/A |

**Table 2.** Summary of compression results for Tucker decomposition and Singular Value Decomposition (SVD). This table highlights trade-offs between compression efficiency, measured by compression ratio and compressed size, and reconstruction fidelity, measured by RMSE and SSIM (for Tucker decomposition only).

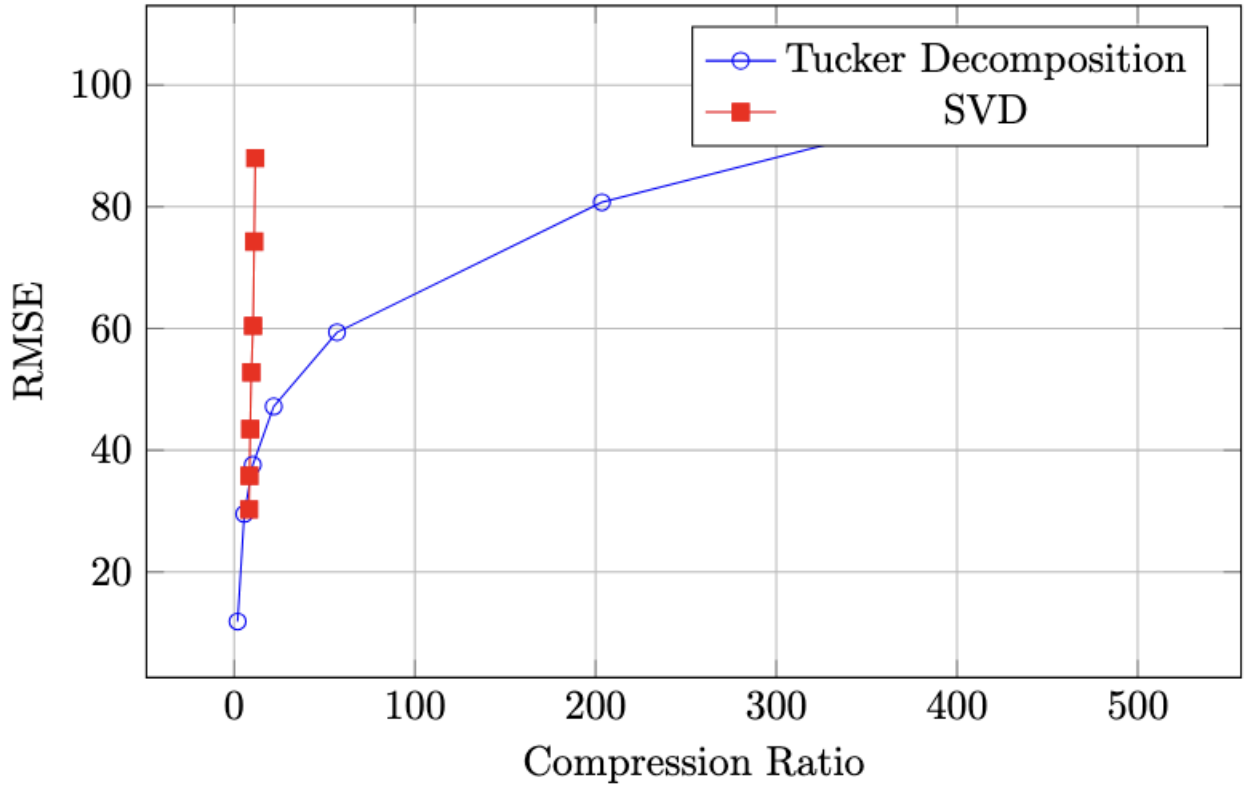

**Figure 5.** This plot illustrates the relationship between compression ratio and reconstruction fidelity, measured by RMSE, for both Tucker decomposition and Singular Value Decomposition (SVD) methods. Tucker decomposition consistently achieves lower RMSE for similar compression ratios compared to SVD, highlighting its ability to better preserve multidimensional relationships in the data. The data points for Tucker decomposition are represented in blue circles, while the SVD results are depicted with red squares.